\documentclass[prd,a4paper,aps,twocolumn,
nofootinbib,nobibnotes,superscriptaddress,preprintnumbers]              {revtex4}

\usepackage[dvips]{graphicx}
\usepackage{amsmath,amssymb,mathrsfs}
\usepackage{bm}
\usepackage{times}
\usepackage{epsfig}
\usepackage{verbatim}
\usepackage{bm}
\usepackage[utf8]{inputenc}
\usepackage{graphics}
\usepackage{graphicx,epsfig,amssymb,amsmath,color, cancel}
\usepackage{soul}
\usepackage[normalem]{ulem}
\usepackage[english]{babel}
\usepackage{amsfonts,amsmath,amssymb,amsthm,mathtools}
\usepackage{comment,enumerate,footnote,graphicx,subfloat,relsize}
\usepackage{array,tabularx,tabu,multirow,framed,afterpage}
\usepackage[usenames,dvipsnames]{xcolor}

\newcommand{\be}{\begin{equation}}
\newcommand{\ee}{\end{equation}}

\begin{document}
\preprint{DESY 18-172}




\title{\Large A light dimuon resonance in $B$ decays?}


\author{\large Filippo Sala\\
\normalsize DESY, Notkestra$\beta$e 85, D-22607 Hamburg, Germany}


\begin{abstract}
The observed deviations from the Standard Model in several $b \to s \mu \mu$ processes can be explained in terms of a new vector boson produced on-shell in $B$ meson decays. A mass of $2.5-3$~GeV and a total width of $10-20\%$ allow to hide the associated dimuon bump in the poorly known charmonium region, and the large invisible decay width can be interpreted in terms of Dark Matter.
This proposal predicts a contribution to the muon anomalous magnetic moment, that could explain the long-standing tension with the Standard Model.
It also predicts sizeable invisible $B$ decays and a peculiar $q^2$-dependence of the lepton flavor universality ratios  $R_K$ and $R_{K^*}$, that could be tested at the LHCb and Belle-II. 
This proceeding is based on~\cite{Sala:2017ihs}, and slightly extends it with comments about Dark Matter.
\end{abstract}

\maketitle

\section{Introduction}
\label{sec:intro}

Processes induced by $b \to s \ell\ell$ transitions arise at one loop in the Standard Model (SM), and in beyond the SM (BSM) theories their size is controlled by parameters that are usually less constrained than those associated to transitions involving lighter quark generations, see e.g.~\cite{Alpigiani:2017lpj}. Therefore they constitute an excellent avenue to look for experimental signs of BSM.

Several deviations from the SM in such processes have been measured in recent years. These include i) a suppression of the lepton flavor universality ratios $R_K = \text{BR}(B^+ \to K^+\mu^+\mu^-)/\text{BR}(B^+ \to K^+e^+e^-)$~\cite{LHCbtalk} and $R_{K^*} = \text{BR}(B \to K^{0*}\mu^+\mu^-)/\text{BR}(B \to K^{0*} e^+e^-)$~\cite{Aaij:2014ora}, both in the dimuon invariant mass bin $q^2 \in [1,6]$~GeV$^2$;
ii) an enhancement of the angular observable $P_5'$ (or $S_5$), defined as in~\cite{Altmannshofer:2008dz,DescotesGenon:2012zf}, in the decays $B^0 \to K^{0*}\mu^+\mu^-$~\cite{Aaij:2013qta,Aaij:2015oid,ATLAS-CONF-2017-023};
iii) a suppression of the branching ratios $B^+\to K^{(*)+} \mu^+ \mu^-$~\cite{Aaij:2014pli}, $B^0 \to K^{(*)0}\mu^+\mu^-$~\cite{Aaij:2014pli,Aaij:2016flj}, $B_s \to\phi \mu^+ \mu^-$~\cite{Aaij:2015esa}.
The tensions in ii) and iii) could partly depend on underestimated theoretical uncertainties, see e.g.~\cite{Ciuchini:2018anp} for recent considerations.
The theoretical cleanliness of the SM prediction for $R_K$ and $R_{K^*}$~\cite{Hiller:2003js,Bordone:2016gaq} makes i) a clearer hint for BSM, of course barring experimental systematics or statistical fluctuations.
As measurements based on new data are expected to be released soon by LHCb, it is a good time to review the BSM explanation of such anomalies put forward in~\cite{Sala:2017ihs}. For convenience of the reader, we start by sketching the idea 
before turning to more quantitative aspects in the following sections.
\medskip

The proposed BSM explanations of the $b \to s \ell\ell$ flavor anomalies typically rely on the introduction of one or more degrees of freedom \textit{heavier} than the $B$ mesons.
Examples include a new vector boson, a leptoquark, or a richer new particle content, see e.g.~\cite{DAmico:2017mtc} for a concise summary with references.
In~\cite{Sala:2017ihs} we have investigated the logical possibility of the anomalies being originated instead from a new particle~\textit{lighter} than the $B$ mesons, produced in $B$ decays, and decaying to muon pairs.\footnote{
Our proposal will not explain the downward fluctuation in the bin $q^2 \in [0.045, 1.1]$~GeV$^2$ of $R_{K^*}$~\cite{Aaij:2017vbb}.
BSM explanations of~\cite{Aaij:2017vbb} involve particles lighter than the muon mass (while our is heavier), and need additional matter content to explain the larger $q^2$ bin of both $R_K$ and $R_{K^*}$, see e.g.~\cite{Ghosh:2017ber,Altmannshofer:2017bsz}.
}
Two immediate challenges to this proposal are a) why haven't we already seen such a resonance as a bump in the invariant mass of two muons coming from some $B$ decay, b) how to decrease the $B$'s branching ratios into muons (as implied by $R_K$ and $R_{K^*}$) rather than the opposite.
These challenges are addressed if a) the new resonance has a mass higher than roughly 2.5~GeV and a large enough width, so the associated dimuon bump can be ``hidden'' in the poorly known region where charmonium resonances like the $J/\psi$ live; b) the new resonance is a vector, so it interferes with the SM contribution and, thanks to the large width and to an appropriate sign choice, it induces a decrease of $R_K$ and $R_{K^*}$ for dimuon invariant masses lower than $m_V^2$.

Not only this proposal turns out to address the anomalies i) to iii) ($P_5'$ actually only partially), but also it predicts a contribution to the muon anomalous magnetic moment that could explain the long standing discrepancy with the SM~\cite{Davier:2010nc}. We will review these aspects in sections~\ref{sec:anomalies} and~\ref{sec:pheno}, together with the observables that will allow to experimentally test this model.
The needed large invisible decay width of the new vector can be intriguingly connected with DM, as we will review in section~\ref{sec:DM}.
From a theoretical point of view, vector bosons of a new broken gauge group are light by symmetry, and they arise in several motivated extensions of the SM. 
Like in~\cite{Sala:2017ihs}, we will not address here the UV origin of the new vector, but just shortly comment about it.

\section{A new light vector and the $b \to s\ell\ell$ anomalies}
\label{sec:anomalies}
We introduce a new massive vector $V$ of mass $m_V$, with the following interactions with the SM in the broken EW phase
\be
\mathcal{L} = 
 \big[ g_{\mu V}\,\bar{\mu} \gamma_\nu \mu
+ g_{\mu A}\,\bar{\mu} \gamma_\nu\gamma_5 \mu
+(g_{bs}\,\bar{s}_L \gamma_\nu b_L + {\rm h.c.})\big] V^\nu\,.
\label{eq:interactions}
\ee
Such couplings may originate, all or part of them, from a remnant of larger flavor symmetries at higher energies and/or from a dark sector portal, possibly in connection with the large invisible $V$ decay width that will be needed. These couplings may also be described in an effective $Z'$ framework~\cite{Fox:2011qd}, where $V$ is the gauge boson of a new group $U(1)'$ under which the SM is neutral, and the SM fields acquire $V$ couplings via mixing with new matter states charged under $U(1)'$.
Note that eq.~(\ref{eq:interactions}) does not need additional sources of flavor violation beyond the SM, indeed the coupling $g_{bs}$ may originate from a $V$ coupling to the top in combination with a SM loop~\cite{Sala:2017ihs,Kamenik:2017tnu}.
As anticipated, we do not comment further on the UV origin of eq.~(\ref{eq:interactions}).

\begin{figure}[t]
\includegraphics[width=\columnwidth]{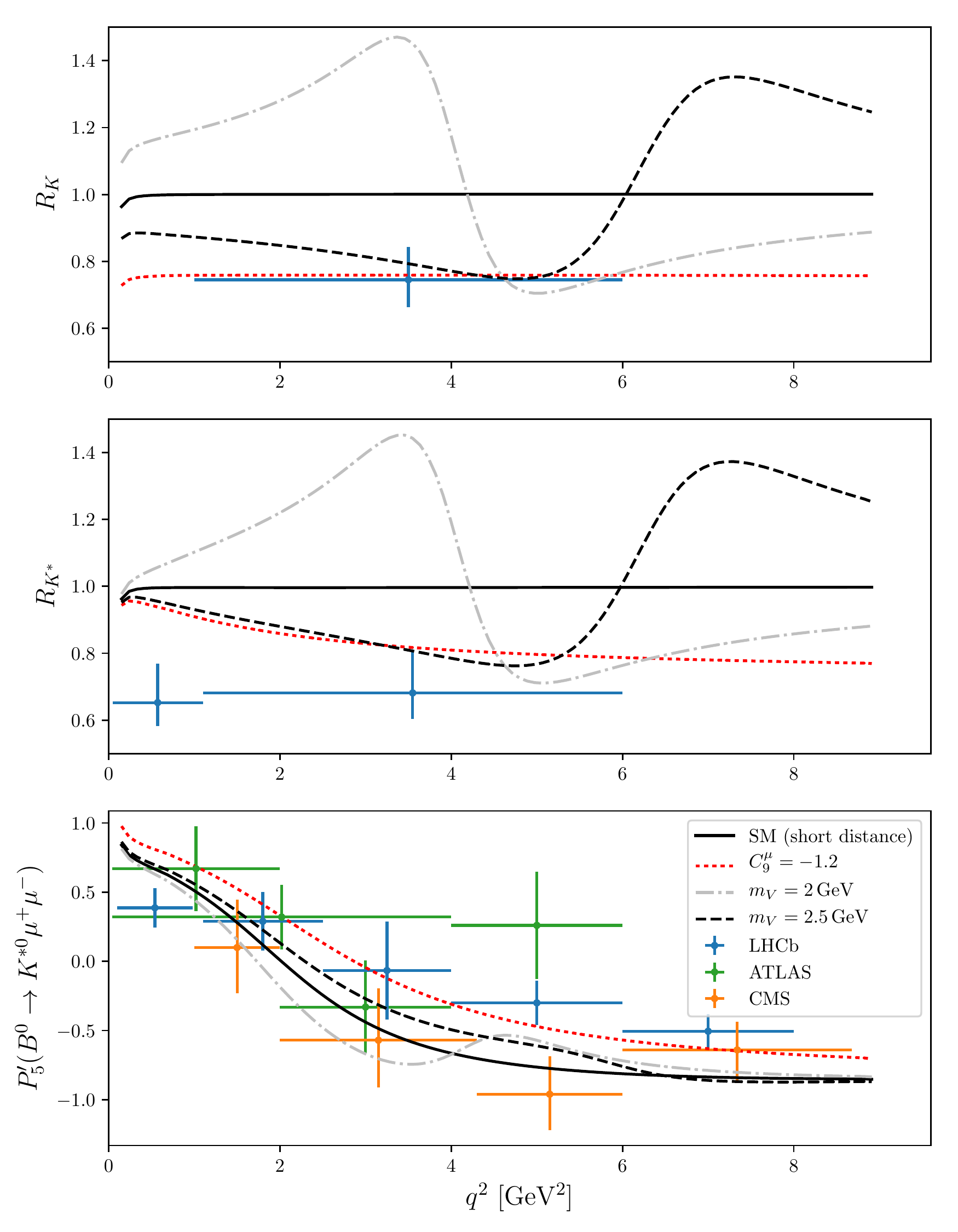}
\caption{$R_K$, $R_{K^*}$, and $P_5'$ as functions of the dimuon invariant mass $q^2$.
We show the SM short distance contribution (continuous black),
a heavy NP contribution to $C_9^\mu$ (dotted red),
and a light vector resonance with 2.5~GeV mass and relative width of 20\%
(dashed black, dot-dashed gray for the case of 2~GeV mass, shown just for comparison with~\cite{Aaij:2016cbx}).
The coupling to muons are fixed as $g_{\mu V} = 0.1$ and $g_{\mu A}\simeq -0.44\,g_{\mu V}$
to explain the muon anomalous magnetic moment, and the one to quarks is fixed to saturate the upper bound from $B$ decays with missing energy eq.~(\ref{eq:invbound}).
Experimental results are shown as dots with error bars.
A marked difference between the light vector and the $C_9^\mu$ contributions
persists
also at values of $q^2$ larger than those  shown here. Figure taken from~\cite{Sala:2017ihs}.}
\label{fig:bins}
\end{figure}

Let us now discuss its impact on the rare semileptonic $B$ decays of interest for the anomalies i)-iii), $B \to K^{(*)} \mu^+ \mu^-$.
To make contact with the usual parametrization of NP in $B$ decays, eq.~(\ref{eq:interactions}) induces the Wilson coefficients
\begin{equation}
C_{9,10}^V = \frac{g_{bs} g_{\mu V,A}/N}{q^2-m_V^2+im_V\Gamma_V},
\label{eq:C9}
\end{equation}
where we have defined $\mathcal{H} = N \sum_i C_i \mathcal{O}_i$,
$\mathcal{O}_9 = (\bar{s}_L \gamma^\nu b_L)(\bar{\mu}\gamma_\nu\mu)$,
$\mathcal{O}_{10} = (\bar{s}_L \gamma^\nu b_L)(\bar{\mu}\gamma_\nu \gamma_5\mu)$ and
$N=2\sqrt{2} G_F V_{tb} V_{ts}^*\alpha/4\pi \simeq -7.7 \times10^{-10}$~GeV$^{-2}$.
For the moment $\Gamma_V$ is a free parameter, that as we now show has to be sizeable.
If the dimuon resonance is fully contained in the bin $q^2 \in [1,6]$~GeV$^2$ relevant for $R_K$ and $R_{K^*}$,
then it would generate an enhancement with respect to the SM, rather than the observed suppression.
If the resonance lies instead at a higher $q^2$, then if it has a large enough width it can induce a suppression of
the SM prediction in the measured bin, because being a vector it can interfere (destructively) with the SM.
A mass larger than $\sqrt{6} \simeq 2.5$~GeV makes the broad bump lie in a region where the SM prediction
for the $q^2$ spectrum is poorly known, because of the sizeable hadronic uncertainties~\cite{Khodjamirian:2010vf,Lyon:2014hpa}
and because the phase of the interference of the SM continuum with the $J/\psi$ resonant contribution is unknown.
A quantitative visualization of the impact of $V$ on $R_K$, $R_K^*$ and $P_5'$
is given in Figure~\ref{fig:bins}. One sees that a width of $\Gamma_V/m_V =20\%$
allows to contain the deviation from the SM for any $q^2 > 6$~GeV$^2$ within 30\% or so
\textit{and} to induce the desired suppression in the bin $q^2 \in [1,6]$~GeV$^2$.
We do not attempt to quantify precisely the maximal tolerated deviation from the SM at $q^2 > 6$~GeV$^2$,
and therefore we refrain from providing a precise upper (lower) limit on the $m_V$ ($\Gamma_V$) that allows to explain
the $B$ anomalies.

The values of the other parameters of eq.~(\ref{eq:interactions}), used in Figure~\ref{fig:bins}, are chosen to improve the
fit in each of $R_K$, $R_K^*$ and $P_5'$ by at least 2$\sigma$ (as determined in~\cite{Sala:2017ihs} using the {\tt flavio}~\cite{flavio} code),
and at the same time to comply with existing constraints.
Before describing such constraints, we observe that they also imply that the large $\Gamma_V$ needed cannot come from eq.~(\ref{eq:interactions}) alone, but it requires the addition of an invisible BSM channel, for example
\be
\mathcal{L}_\text{inv} = g_\chi\,\bar{\chi}\gamma_\nu \chi\, V^\nu \quad \Rightarrow \quad \Gamma_V \simeq \frac{m_V}{12 \pi}\, g_\chi^2\,,
\label{eq:inv}
\ee
with $\chi$ invisible for LHC detectors and $g_\chi \gtrsim 2$.

\section{Constraints and phenomenological implications}
\label{sec:pheno}

\begin{figure*}[t]
\includegraphics[width=\textwidth]{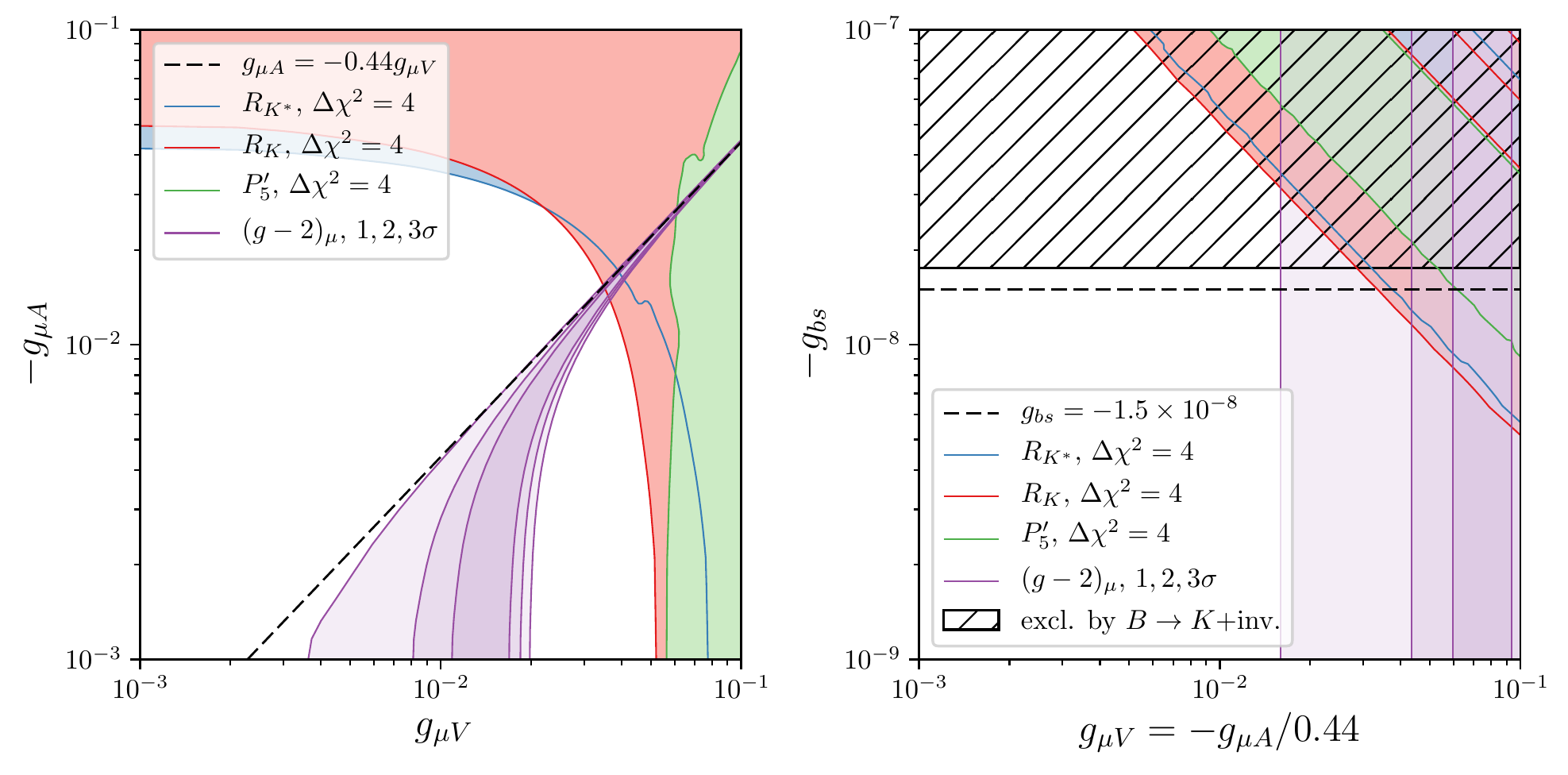}
\caption{Regions preferred by the $b \to s \ell \ell$ anomalies and by other constraints for the
benchmark value of $m_V=2.5$~GeV.
Left: $g_{\mu V}$ vs. $g_{\mu A}$, with $g_{bs}=-1.5\times10^{-8}$.
For a visualization on this plane of constraints from the $Z$ boson dimuon line-shape,
that challenge the explanation of $P_5'$ but not that of $R_K$ and $R_{K^*}$, see~\cite{Bishara:2017pje}.
Right: $g_{\mu V}$ vs. $g_{bs}$, with $g_{\mu A}=-0.44 g_{\mu V}$.
Figure taken from~\cite{Sala:2017ihs}.
}
\label{fig:parameters}
\end{figure*}
Let us now describe some main phenomenological implications of eqs.~(\ref{eq:interactions}) and (\ref{eq:inv}),
starting from the $V$ couplings to muons $g_{\mu V,A}$.
A loop containing $V$ induces an anomalous magnetic moment for the muon
\begin{equation}
\delta a_\mu =
\frac{m_\mu^2}{m_V^2}
\frac{g_{\mu V}^2\,f_V \Big( \frac{m_\mu^2}{m_V^2}\Big) -5 g_{\mu A}^2\,f_A \Big( \frac{m_\mu^2}{m_V^2}\Big)}{12\pi^2},
\label{eq:g-2}
\end{equation}
where $f_{V,A} = 1$ up to subleading orders in $m_\mu^2/m_V^2$.~\footnote{%
For the full expressions of the loop functions see e.g.~\cite{Kannike:2011ng}, in that notation $f_V(x) = \frac{3}{8}(2\,F_Z(x)+G_Z(x))$,
$f_A(x) = \frac{3}{40}(G_Z(x)-2\,F_Z(x))$.}
It is interesting that, for $|g_{\mu A}|<\sqrt{5}|g_{\mu V}|$, the contribution in eq.~(\ref{eq:g-2}) has the right sign to resolve  $(g-2)_\mu$ anomaly~\cite{Davier:2010nc}
\begin{equation}
a_\mu^\text{exp} - a_\mu^\text{SM} = (287\pm 80) \times 10^{-11}.
\end{equation}
For $m_V\gg m_\mu$ and fixing for definiteness $g_{\mu A}=0$, requiring the contribution not to overshoot the measurement by more than
$3\sigma$ implies $g_{\mu}/[m_V/\text{GeV}]\lesssim 0.008$. 
In principle the bound can be weakened or even removed by tuning $g_{\mu V}$ vs. $g_{\mu A}$, and this possibility will be crucial in allowing to explain the flavor anomalies in $b \to s \ell\ell$ decays\footnote{Explanations of the $b \to s \ell\ell$ anomalies in terms of heavier vectors, instead, allow to explain also $(g-2)_\mu$ only at the price of adding extra matter, see e.g.~\cite{Belanger:2015nma,Allanach:2015gkd,DiChiara:2017cjq}.
}.
The first two terms in eq.~(\ref{eq:interactions}) 
are also somehow severely constrained by
the precise measurements of the invariant mass of dimuons from Drell-Yan production, that as shown in~\cite{Bishara:2017pje}
reduce the parameter space where $P_5'$ can be explained,
but leave ample margin to address $R_K$ and $R_{K^*}$.


Let us now turn to the coupling $g_{bs}$.
The most important constraint on it
arises from the combination with the invisible decay width, 
and comes from $B\to K\chi\chi$.
This process experimentally gives the
same signature of the SM process $B \to K\nu \bar{\nu}$,
which has been looked for at both Belle and Babar~\cite{Grygier:2017tzo,Lutz:2013ftz,Lees:2013kla,delAmoSanchez:2010bk},
A combination of those measurements gives~$\text{BR}(B\to K \text{inv.}) <1.5\times 10^{-5}$ at 95\%CL~\cite{Sala:2017ihs}.
The SM prediction for $\text{BR}(B\to K\nu \bar\nu)$ is given by $0.5\times 10^{-5}$~\cite{Buras:2014fpa}, with a relative error
of order 10\%.
We therefore impose the upper limit $\text{BR}(B \to K \chi \chi) < 10^{-5} $.
In our model, in the narrow width approximation for $V$ and using $\text{BR}(V\to\chi\chi) \simeq 1$ we have
\begin{equation}
\text{BR}(B\to K\chi\chi)\simeq \frac{g_{bs}^2}{64\pi} \frac{m_B^3}{\Gamma_B m_V^2}~ \lambda^{3/2} ~[f_+(m_V^2)]^2\,,
\end{equation}
where $\Gamma_B$ is the $B$ meson total width,
$\lambda=1+ \hat{m}_V^4 + \hat{m}_K^4 - 2(\hat{m}_K^2 + \hat{m}_V^2 + \hat{m}_K^2 \hat{m}_V^2)$ with $\hat{m}_i=m_i/m_B$,
and $f_+(q^2) $ is a form factor that we take from~\cite{Bailey:2015dka}.
The resulting upper bound on $g_{bs}$ reads
\begin{equation}
g_{bs} \lesssim 0.7\times10^{-8} \times (m_V/\text{GeV}) \,.
\label{eq:invbound}
\end{equation}
This bound becomes weaker if $\Gamma_V$ is large enough to invalidate the narrow width approximation.

Constraints from modifications of the $Z\mu\mu$ coupling, $B_s-\bar{B}_s$ mixing and $B_s \to \mu^+\mu^-$
turn out to be less relevant.
Constraints could also arise from $Z \to 4\mu$ and $Z \to 2\mu$+inv., but to our knowledge the
first has been performed only down to $m_{\mu\mu} = 4$~GeV, and the second has not been performed.
We refer the reader to~\cite{Sala:2017ihs} for a discussion of these constraints.

To conclude, we find that an explanation of the $b \to s \ell\ell$ anomalies and $(g-2)_\mu$ is possible,
compatibly with all the other constraints, for
\begin{equation}
10^{-9} \lesssim |g_{bs}\,g_{\mu V}| \lesssim 3\times 10^{-9}\,,
\end{equation}
where we have fixed $g_{\mu A} \simeq -0.44 g_{\mu V}$ as a benchmark motivated by $(g-2)_\mu$.
The region where the three sets of flavour observables,
$R_K$, $R_{K^*}$ and $P_5'$, are improved by more than $2\sigma$ each is shown in Figure~\ref{fig:parameters},
together with the region preferred by $(g-2)_{\mu}$, and with the constraint $B \to K+$invisible.
The other possible relative sign between $g_{\mu A}$ and $g_{\mu V}$ does not change
the prediction for $(g-2)_\mu$, but it induces a slightly worse fit
of the $b \to s \ell \ell$ anomalies and is not shown.

\section{Connection with Dark Matter}
\label{sec:DM}

Could the new particle(s) to which $V$ decays invisibly constitute the observed Dark Matter (DM),
for the same values of the parameters that solve the flavour tensions?
For simplicity let us discuss the case of one new stable particle.
Its large coupling to the new vector $V$ and the sizeable values of $g_{\mu V,A}$, demanded to explain the flavour anomalies, imply that this particle was in equilibrium with the SM in the Early Universe.
The key quantity to investigate then becomes the annihilation cross section of a DM pair into SM particles.

When the DM is heavier than a muon the dominant annihilation is into muon pairs.
The related averaged cross section times relative velocity reads
\begin{equation}
\langle \sigma v \rangle
\simeq \frac{g_\chi^2}{\pi}\,\frac{g_{\mu V}^2 \big(m_\chi^2 + \frac{m_\mu^2}{2}\big)
+ g_{\mu A}^2 \big(m_\chi^2 - m_\mu^2\big)}{(m_V^2 - 4 m_\chi^2)^2}\sqrt{1-\frac{m_\mu^2}{m_\chi^2}},
\label{eq:DMfermion}
\end{equation}
in the case of fermion DM $\chi$, and
\begin{equation}
\langle \sigma v \rangle \simeq
v^2 \frac{g_\phi^2}{3\pi} \frac{g_{\mu V}^2 \big(m_\phi^2 + \frac{m_\mu^2}{2}\big)
+g_{\mu A}^2 \big(m_\phi^2 - m_\mu^2\big)}{(m_V^2 - 4 m_\phi^2)^2} 
\sqrt{1-\frac{m_\mu^2}{m_\phi^2}}.
\label{eq:DMscalar}
\end{equation}
in the case of scalar DM $\phi$ with coupling $i\, g_\phi V^\nu \phi^*\partial_\nu \phi + {\rm h.c.}$.
For the preferred values of the invisible decay width of $V$ and of $g_{\mu V,A}$, these expressions yield to $\langle \sigma v \rangle$ in the ballpark of $10^{-(22-21)} $cm$^3$/sec for fermion DM, and of $10^{-(24-23)} $cm$^3$/sec at freeze-out for scalar DM (note here the $p$-wave suppression).
While the annihilation of $\chi$ with $\bar{\chi}$ (or of $\phi$ with $\phi^*$) is too large to obtain the observed DM abundance via thermal freeze-out, the efficient depletion of the symmetric DM population makes this a viable DM model in presence of an appropriate initial asymmetry.

If the particle to which $V$ decays invisibly is instead lighter than a muon, then it would reproduce the DM abundance via thermal freeze-out if, additionaly, it has couplings to electrons $g_{eV,A}$ of the order of $10^{-1}-10^{-2}$ times the favoured couplings to muons, e.g. in the fermion $\chi$ case. Note that such couplings to electrons would not spoil the explanation of the flavor anomalies.
Without relying on such extra coupling to electrons, the leading annihilating channels would be either $\nu\bar{\nu}$ or $\bar{\mu} e \nu \bar{\nu}$. The related annihilation cross sections would be suppressed, with respect to those in eqs~(\ref{eq:DMfermion}) and (\ref{eq:DMscalar}), by an extra factor of $\sim (m_\chi/m_W)^2 \times (\alpha_2/4\pi)^2 <(<) 10^{-10}$, thus leading to a bad overclosure of the Universe.

We conclude this section by noting that the self-interacting cross section of such DM candidate, $\sigma/m_\chi \sim (g_\chi^4/8\pi)\times m_\chi/m_V^4$, is much smaller than the value of $\approx$ barn/GeV needed for it to be observable in the distribution of matter in galaxies and clusters (and therefore also too small to solve the related putative anomalies in small scale structures), see e.g.~\cite{Tulin:2017ara} for a review. An analogous conclusion holds for the scalar DM candidate.
\medskip

\section{Conclusions}
\label{sec:conclusions}
In this proceeding we have reviewed the explanation of the neutral current $B$ anomalies and of the muon anomalous magnetic moment proposed in~\cite{Sala:2017ihs}, in terms of a new vector dimuon resonance with a mass $\gtrsim 2.5$ GeV and a sizeable invisible decay width.
The latter can be intriguingly connected to cold Dark Matter candidates with various cosmological histories.

Independently of Dark Matter, this scenario predicts other deviations from the SM in $B$ decays, most notably
\begin{itemize}
\item[$\diamond$] a strong $q^2$ dependence of the lepton flavor universality ratios $R_K$ and $R_{K^*}$, unique to this explanation of the $b \to s \ell \ell$ anomalies;
\item[$\diamond$] a branching ratio of $B \to K$~invisible close to its present upper limit.
\end{itemize}
Measurements of these observables will allow to confirm or rule out this possibility in the near future at the LHCb and Belle-II.

This proposal addressed an unexplored logical possibility to explain the $B$ anomalies, having demonstrated its phenomenological viability the next step would be to include it in a more complete UV picture.
We do not perform this step here, but we mention that another constraint that would immediately become important is the one from neutrino trident production~\cite{Altmannshofer:2014pba}, that would rule out the muon couplings that explain the flavor anomalies if $g_{\nu_L} = g_{\mu_L} = g_{\mu V} - g_{\mu A}$.

\section*{Acknowledgements}
I thank David Straub for the pleasant and efficient collaboration on the topics presented in this proceeding, and for comments on the manuscript.
I am grateful to the organisers of ``FPCapri2018'' and ``The Future of BSM Physics'' for the lovely conference and workshop, and MITP for partial support during the two weeks in Capri. I acknowledge partial support by a PIER Seed Project funding (Project ID PIF-2017-72).




\bibliography{FSala_LightRK}







\end{document}